\documentclass{article}%
\usepackage{amsfonts}
\usepackage{amssymb}
\usepackage{amsmath}%
\usepackage{graphicx}
\providecommand{\U}[1]{\protect\rule{.1in}{.1in}}
\newtheorem{theorem}{Theorem}

\newtheorem{lemma}[theorem]{Lemma}

\newtheorem{remark}[theorem]{Remark}

\newenvironment{proof}[1][Proof]{\noindent\textbf{#1.} }{\ \rule{0.5em}{0.5em}}
\begin{document}

\title{Entanglement cost and distillable entanglement of symmetric states}
\author{Keiji Matsumoto\\National Institute of Informatics, 2-1-2, Hitotsubashi, Chiyoda-ku, \\Tokyo 101-8430,, Japan \\Sorst Quantum Computation and Information project, JST,\\Daini Hongo White Bldg.201, 5-28-3,Hongo, Bunkyo-ku, \\Tokyo 113-0033, Japan\\ }
\maketitle

\begin{abstract}
We compute entanglement cost and distillable entanglement of states supported
on symmetric subspace. Not only giving general formula, we apply them to the
output states of optimal cloning machines. Surprisingly, under some settings,
the optimal $n$ to $m$ \ clone and true $m$ copies are the same in
entanglement measures. However, they differ in the error exponent of
entanglement dilution. We also presented a general theory of entanglement
dilution which is applicable to any non-i.i.d sequence of states.

\end{abstract}

\section{Introduction}

In asymptotic theory of entanglement,it is often assumed that the given state
is in the form of $\rho^{\otimes n}$, or independent identical copies of a
state $\rho$ (i.i.d. ensemble, hereafter). In some important cases, however,
this assumption is not necessarily true. For example, in study of local
copying \cite{ACP}\cite{OwariHayashi:2006}, we have to treat the optimal clone
of a bipartite state: Given $n$ copies of them, its optimal $n$ to $m$ clone
is not close to i.i.d ensemble at all. The purpose of this manuscript is to
give explicit, tractable formula of entanglement cost and distillable
entanglement of non-i.i.d. states.

In the manuscript, we discuss entanglement cost of general non-i.i.d. state,
using information spectrum \cite{Han}\cite{HanBook}\cite{Hayashi:2006}%
\cite{HayashiNagaoka:2003}. (in quantum information jargon, it is called
smooth Reny entropy.) This formula, however, contains maximization which
cannot be solved in most of the cases.

Therefore, second, we present a formula without maximization for symmetric
states, or states supported on symmetric subspace. For such states, we also
give distillable entanglement, too. Remarkably, the optimal entangle
distillation is possible without knowing the input state except the fact that
it is a symmetric state. This is a generalization of universal entanglement
concentration in \cite{MatsumotoHayashi}.

Finally, we apply this theory to output states of $n$ to $m$ cloning machines.
We assume that the input is $n$ copies of the identical pure states, and $m$
equals $rn$ for some constant $r$. The following two kinds of cloning machines
are considered.

First example is the machine optimal for the case where the Schmidt basis of
the input is known except its phases. To our surprise, both the entanglement
cost and distillable entanglement are $m$ times entropy of entanglement of the
input state. Hence, optimal clone and real $m$ copies are the same in its
entanglement measures. We also computed error exponent of entanglement
distillation and dilution, and showed they are worse than real $m$ copies.

Second example is the machine optimal for all the possible pure bipartite
states. For this, we only proved that $m$ times the entropy of entanglement
cost of the input state is an upperbound of the entanglement cost. Our
conjecture is that this upperbound is the entanglement cost and, at the same
time, the distillable entanglement.

\section{A general theory of entanglement cost}

Below, $\left\vert \Phi_{D}\right\rangle $ is a maximally entangled state with
Schmidt rank $D$, and $\mathrm{F}\left(  \rho,\left\vert \Phi_{D}\right\rangle
\right)  $ is the optimal fidelity of generating $\rho$ from $\left\vert
\Phi_{D}\right\rangle $ by LOCC.

\begin{lemma}%
\[
\mathrm{F}^{D}\left(  \rho\right)  =\max_{\{q_{i},\left\vert \phi
_{i}\right\rangle \}}\sum_{i}\sum_{j=1}^{D}q_{i}p_{j}^{\phi_{i}},
\]
where $p_{j}^{\phi_{i}}$ is the $j$th largest Schmidt coefficient of
$\left\vert \phi_{i}\right\rangle $, and the maximization is taken over pure
state ensembles with $\sum_{i}q_{i}\left\vert \phi_{i}\right\rangle
\left\langle \phi_{i}\right\vert =\rho$.
\end{lemma}

\begin{proof}
Observe%
\begin{align*}
\mathrm{F}^{D}\left(  \rho\right)   &  =\max_{\{A_{i}\}}\mathrm{F}\left(
\rho,\sum_{j}A_{j}\left\vert \Phi_{D}\right\rangle \left\langle \Phi
_{D}\right\vert A_{j}^{\dagger}\right)  \\
&  =\max_{\{A_{i}\}}\max_{\{q_{i},\left\vert \phi_{i}\right\rangle
\}}\left\vert \sum_{i,j}\sqrt{q_{i}}\left\langle \phi_{i}\right\vert
A_{i}\left\vert \Phi_{D}\right\rangle \right\vert ^{2}.
\end{align*}
We solve maximization over $\left\{  A_{i}\right\}  $. Since they are LOCC,
the Schmidt rank of $A_{i}\left\vert \Phi_{D}\right\rangle $ cannot be more
than $D$. Therefore, it is optimal if
\[
A_{i}\left\vert \Phi_{D}\right\rangle =\frac{c_{i}}{\sqrt{\sum_{j=1}^{D}%
p_{j}^{\phi_{i}}}}\sum_{j=1}^{D}\sqrt{p_{j}^{\phi_{i}}}\left\vert
j\right\rangle \left\vert j\right\rangle ,
\]
and this is possible for any $\left\{  c_{i}\right\}  $ with $\sum
_{i}\left\vert c_{i}\right\vert ^{2}=1$. Therefore,%
\begin{align*}
\mathrm{F}^{D}\left(  \rho\right)   &  =\max_{\{q_{i},\left\vert \phi
_{i}\right\rangle \}}\max_{\left\{  c_{i}\right\}  :\sum_{i}\left\vert
c_{i}\right\vert ^{2}=1}\left\vert \sum_{i}\sqrt{q_{i}}\frac{c_{i}}{\sqrt
{\sum_{j=1}^{D}p_{j}^{\phi_{i}}}}\sum_{j=1}^{D}p_{j}^{\phi_{i}}\right\vert
^{2}\\
&  =\max_{\{q_{i},\left\vert \phi_{i}\right\rangle \}}\max_{\left\{
c_{i}\right\}  :\sum_{i}\left\vert c_{i}\right\vert ^{2}=1}\left(  \sum
_{i}c_{i}\sqrt{q_{i}\sum_{j=1}^{D}p_{j}^{\phi_{i}}}\right)  ^{2}\\
&  =\max_{\{q_{i},\left\vert \phi_{i}\right\rangle \}}\sum_{i}q_{i}\sum
_{j=1}^{D}p_{j}^{\phi_{i}}.
\end{align*}

\end{proof}

\bigskip

Given a sequence $\left\{  \rho^{n}\right\}  _{n=1}^{\infty}$ of bipartite
quantum states, we consider a sequence of purestate ensembles $\{q_{i}%
^{n},\left\vert \phi_{i}^{n}\right\rangle \}$ with $\sum_{i}q_{i}%
^{n}\left\vert \phi_{i}^{n}\right\rangle \left\langle \phi_{i}^{n}\right\vert
=\rho^{n}$. \ Let $p_{j}^{n,i}$, where $j$ runs from $1$ to the Schmidt rank
of $\left\vert \phi_{i}^{n}\right\rangle $, be the Schmidt coefficients of
$\left\vert \phi_{i}^{n}\right\rangle $ ($p_{1}^{n,i}\geq p_{2}^{n,i}\cdots$).
Then, $q_{i}^{n}p_{j}^{n,i}$ defines a probability distribution over $\left(
i,j\right)  $. At the same time, the value $p_{j}^{n,i}$ can be viewed as a
random variable, where $\left(  i,j\right)  $ occurs with the probability
$q_{i}^{n}p_{j}^{n,i}$.

Given a sequence of probability distributions $\left\{  P^{n}\right\}
_{n=1}^{\infty}$ over some discrete set, we define a notion of probabilistic
limsup of a random variable $X^{n}$, denote by $\mathrm{p-}\varlimsup
_{n\rightarrow\infty}X^{n}$, the minimum of $x$ with%

\[
\lim_{n\rightarrow\infty}P^{n}\left\{  \,i\,\,;\,\,X^{n}\leq x\right\}  =1.
\]
We also denote by $\mathrm{p-}\varliminf_{n\rightarrow\infty}X^{n}$, the
maximum of $x$ with%

\[
\lim_{n\rightarrow\infty}P^{n}\left\{  \,i\,\,;\,\,X^{n}\geq x\right\}  =1.
\]

\begin{theorem}
Given a sequence $\left\{  \rho^{n}\right\}  _{n=1}^{\infty}$ of bipartite
quantum states, we have
\[
E_{c}\left(  \left\{  \rho^{n}\right\}  _{n=1}^{\infty}\right)  =\inf
_{\{q_{i}^{n},\left\vert \phi_{i}^{n}\right\rangle \}}\mathrm{p-}%
\varlimsup_{n\rightarrow\infty}\frac{-1}{n}\log p_{j}^{n,i}%
\]
where \ $\mathrm{p}-\overline{\lim}_{n\rightarrow\infty}$ is with respect to
$\left\{  q_{i}^{n}p_{j}^{n,i}\right\}  _{n=1}^{\infty}$ , and infimum is
taken over all the sequences of pure state ensembles $\{q_{i}^{n},\left\vert
\phi_{i}^{n}\right\rangle \}$ with $\sum_{i}q_{i}^{n}\left\vert \phi_{i}%
^{n}\right\rangle \left\langle \phi_{i}^{n}\right\vert =\rho^{n}$.
\end{theorem}

\begin{proof}
\bigskip We use the technique which repeatedly used in \cite{Han}. "$\leq$" is
proved as follows. For any $j_{0}$, we have
\[
1\geq\sum_{j=1}^{j_{0}}p_{j}^{n,i}\geq j_{0}p_{j_{0}}^{n,i},
\]
implying
\[
\left\{  j:p_{j}^{n,i}\geq c^{-1}\right\}  \subset\left\{  j:j\leq c\right\}
,
\]
and
\begin{equation}
\mathfrak{C}_{R}^{n}\subset\mathfrak{D}_{R}^{n},\label{jPj<1}%
\end{equation}
where
\begin{align*}
\mathfrak{C}_{R}^{n} &  :=\left\{  \left(  i,j\right)  ;\,\,p_{j}^{n,i}%
\geq2^{-nR}\right\}  ,\\
\mathfrak{D}_{R}^{n} &  :=\left\{  \left(  i,j\right)  ;\,\,j\leq
2^{nR}\right\}  .
\end{align*}
Therefore,
\begin{equation}
\sum_{\left(  i,j\right)  \in\mathfrak{C}_{R}^{n}}q_{i}^{n}p_{j}^{n,i}\leq
\sum_{\left(  i,j\right)  \in\mathfrak{D}_{R}^{n}}q_{i}^{n}p_{j}^{n,i}%
\leq\mathrm{F}^{2^{nR}}\left(  \rho^{\otimes n}\right)
.\label{fidelity-lower}%
\end{equation}
If $R>\mathrm{p-}\varlimsup_{n\rightarrow\infty}\frac{-1}{n}\log p_{j}^{n,i}$,
the left most side tends to $1$ as $n\rightarrow\infty$, meaning that
\[
E_{c}\left(  \left\{  \rho^{n}\right\}  _{n=1}^{\infty}\right)  \leq
\mathrm{p-}\varlimsup_{n\rightarrow\infty}\frac{-1}{n}\log p_{j}^{n,i}%
\]
holds for any pure state ensembles $\{q_{i}^{n},\left\vert \phi_{i}%
^{n}\right\rangle \}$ with $\sum_{i}q_{i}^{n}\left\vert \phi_{i}%
^{n}\right\rangle \left\langle \phi_{i}^{n}\right\vert =\rho^{n}$, and we have
"$\leq$". "$\geq$" is proved as follows. Since%
\[
\overline{\mathfrak{C}_{R+\gamma}^{n}}\subset\left(  \overline{\mathfrak{C}%
_{R+\gamma}^{n}}\cap\mathfrak{D}_{R}^{n}\right)  \cup\overline{\mathfrak{D}%
_{R}^{n}},
\]
we have
\[
\sum_{\left(  i,j\right)  \in\overline{\,\mathfrak{C}_{R+\gamma}^{n}}}%
q_{i}^{n}p_{j}^{n,i}\leq\sum_{\left(  i,j\right)  \in\overline{\mathfrak{C}%
_{R+\gamma}^{n}}\cap\,\mathfrak{D}_{R}^{n}}q_{i}^{n}p_{j}^{n,i}+\sum_{\left(
i,j\right)  \in\overline{\mathfrak{D}_{R}^{n}}}q_{i}^{n}p_{j}^{n,i}.
\]
Since $\left\vert \mathfrak{D}_{R}^{n}\right\vert =2^{nR}$,
\begin{align*}
\sum_{\left(  i,j\right)  \in\overline{\mathfrak{C}_{R+\gamma}^{n}}%
\cap\mathfrak{D}_{R}^{n}}q_{i}^{n}p_{j}^{n,i} &  \leq\sum_{\left(  i,j\right)
\in\mathfrak{D}_{R}^{n}}q_{i}^{n}2^{-n\left(  R+\gamma\right)  }\\
&  \leq\sum_{i}q_{i}^{n}2^{nR}\cdot2^{-n\left(  R+\gamma\right)  }\\
&  =2^{-n\gamma}.
\end{align*}
Hence, with a proper choice of pure state ensemble $\left\{  q_{i}%
^{n},\,\left\vert \phi_{i}^{n}\right\rangle \right\}  $, for any $\epsilon
>0$,
\begin{align}
1-\mathrm{F}^{2^{nR}}\left(  \rho^{n}\right)  +\epsilon &  \geq1-\sum_{\left(
i,j\right)  \in\mathfrak{D}_{R}^{n}}q_{i}^{n}p_{j}^{n,i}\nonumber\\
&  =\sum_{\left(  i,j\right)  \in\overline{\mathfrak{D}_{R}^{n}}}q_{i}%
^{n}p_{j}^{n,i}\nonumber\\
&  \geq\sum_{\left(  i,j\right)  \in\overline{\,\mathfrak{C}_{R+\gamma}^{n}}%
}q_{i}^{n}p_{j}^{n,i}-\sum_{\left(  i,j\right)  \in\overline{\mathfrak{C}%
_{R+\gamma}^{n}}\cap\mathfrak{D}_{R}^{n}}q_{i}^{n}p_{j}^{n,i}\nonumber\\
&  \geq\sum_{\left(  i,j\right)  \in\overline{\,\mathfrak{C}_{R+\gamma}^{n}}%
}q_{i}^{n}p_{j}^{n,i}-2^{-n\gamma}.\label{fidelity-upper}%
\end{align}
Suppose
\begin{align*}
R &  <\inf_{\left[  \left\{  q_{i}^{n},\,\left\vert \phi_{i}^{n}\right\rangle
\right\}  \right]  _{n=1}^{\infty}}\mathrm{p-}\varlimsup_{n\rightarrow\infty
}\frac{-1}{n}\log p_{j}^{n,i},\\
&  \leq\mathrm{p-}\varlimsup_{n\rightarrow\infty}\frac{-1}{n}\log p_{j}^{n,i}.
\end{align*}
Then, we can choose $\gamma>0$ with there is $R+\gamma<\mathrm{p-}%
\varlimsup_{n\rightarrow\infty}\frac{-1}{n}\log p_{j}^{n,i}$, so that the last
end of this inequality does not vanish as $n\rightarrow\infty$. Hence, we
cannot do entanglement dilution with high fidelity, if $R<\mathrm{p-}%
\varlimsup_{n\rightarrow\infty}\frac{-1}{n}\log p_{j}^{n,i}$. Therefore, we
have "$\geq$".
\end{proof}

\begin{theorem}
Suppose
\[
R<\inf_{\left[  \left\{  q_{i}^{n},\,\left\vert \phi_{i}^{n}\right\rangle
\right\}  \right]  _{n=1}^{\infty}}\mathrm{p-}\varliminf_{n\rightarrow\infty
}\frac{-1}{n}\log p_{j}^{n,i},
\]
where \ $\mathrm{p-}\varliminf_{n\rightarrow\infty}$ is with respect to
$\left\{  q_{i}^{n}p_{j}^{n,i}\right\}  _{n=1}^{\infty}$ , and infimum is
taken over all the sequences of pure state ensembles $\{q_{i}^{n},\left\vert
\phi_{i}^{n}\right\rangle \}$ with $\sum_{i}q_{i}^{n}\left\vert \phi_{i}%
^{n}\right\rangle \left\langle \phi_{i}^{n}\right\vert =\rho^{n}$. Then,
\[
\mathrm{F}^{2^{nR}}\left(  \rho^{n}\right)  \rightarrow0.
\]
Also, if
\[
R>\inf_{\left[  \left\{  q_{i}^{n},\,\left\vert \phi_{i}^{n}\right\rangle
\right\}  \right]  _{n=1}^{\infty}}\mathrm{p-}\varliminf_{n\rightarrow\infty
}\frac{-1}{n}\log p_{j}^{n,i},
\]%
\[
\varlimsup_{n\rightarrow\infty}\mathrm{F}^{2^{nR}}\left(  \rho^{n}\right)  >0.
\]

\end{theorem}

\begin{proof}
We use the inequality (\ref{fidelity-upper}). With a proper choice of pure
state ensemble $\left\{  q_{i}^{n},\,\left\vert \phi_{i}^{n}\right\rangle
\right\}  $, for any $\epsilon>0$,%
\[
1-\mathrm{F}^{2^{nR}}\left(  \rho^{n}\right)  +\epsilon\geq\sum_{\left(
i,j\right)  \in\overline{\,\mathfrak{C}_{R+\gamma}^{n}}}q_{i}^{n}p_{j}%
^{n,i}-2^{-n\gamma}.
\]
Choose $\gamma$ with
\begin{align*}
R+\gamma &  <\inf_{\{q_{i}^{n},\left\vert \phi_{i}^{n}\right\rangle
\}}\mathrm{p-}\varliminf_{n\rightarrow\infty}\frac{-1}{n}\log p_{j}^{n,i}\\
&  \leq\mathrm{p-}\varliminf_{n\rightarrow\infty}\frac{-1}{n}\log p_{j}^{n,i}.
\end{align*}
Then, due the definition of probabilistic liminf,%
\[
\lim_{n\rightarrow\infty}\sum_{\left(  i,j\right)  \in\overline{\,\mathfrak{C}%
_{R+\gamma}^{n}}}q_{i}^{n}p_{j}^{n,i}=1.
\]
Therefore,%
\begin{align*}
\varlimsup_{n\rightarrow\infty}\mathrm{F}^{2^{nR}}\left(  \rho^{\otimes
n}\right)   &  \leq\lim_{n\rightarrow\infty}2^{-n\gamma}+\epsilon\\
&  =\epsilon.
\end{align*}
Since $\epsilon$ is arbitrary positive number, our first assertion is proved.
Next, due to (\ref{fidelity-lower}), we have%
\[
\sum_{\left(  i,j\right)  \in\mathfrak{C}_{R}^{n}}q_{i}^{n}p_{j}^{n,i}%
\leq\mathrm{F}^{2^{nR}}\left(  \rho^{n}\right)  .
\]
Due to the definition of probabilistic liminf, if $R>\inf_{\left[  \left\{
q_{i}^{n},\,\left\vert \phi_{i}^{n}\right\rangle \right\}  \right]
_{n=1}^{\infty}}\mathrm{p-}\varliminf_{n\rightarrow\infty}\frac{-1}{n}\log
p_{j}^{n,i}$, liminf of the left hand side does not vanish. Hence, we have our
second assertion.
\end{proof}

\section{A standard form of symmetric states}

Below, we discuss entanglement of symmetric states, or states supported on the
symmetric subspace of $\left(  \mathcal{H}_{A}\otimes\mathcal{H}_{B}\right)
^{\otimes n}$, where $\mathcal{H}_{A}\simeq\mathcal{H}_{B}\simeq\mathcal{H}$
and $\dim\mathcal{H}=d$. \ For that purpose, we introduce a standard form of
such states in this section. It suffices to give a standard form for a pure
symmetric state, since a mixed state is convex combination of them.

Suppose we are given $n$-copies of unknown pure bipartite state $\left\vert
\phi\right\rangle \in\mathcal{H}_{A}\otimes\mathcal{H}_{B}$, which is unknown.
Here we assume $\mathcal{H}_{A}\simeq\mathcal{H}_{B}\simeq\mathcal{H}$ and
$\dim\mathcal{H}=d$.

It is known that $\left\vert \phi\right\rangle ^{\otimes n}$, where
$\left\vert \phi\right\rangle \in\mathcal{H}_{A}\otimes\mathcal{H}_{B}$, has
the standard form defined as follows. Note $|\phi\rangle^{\otimes n}$ is
invariant by the reordering of copies, or the action of the permutation
$\sigma$ in the set $\{1,\ldots n\}$ such that
\begin{equation}
\bigotimes_{i=1}^{n}|h_{i,A}\rangle|h_{i,B}\rangle\mapsto\bigotimes_{i=1}%
^{n}|h_{\sigma^{-1}(i),A}\rangle|h_{\sigma^{-1}(i),B}\rangle,\label{sym}%
\end{equation}
where $|h_{i,A}\rangle\in\mathcal{H}_{A}$ and $|h_{i,B}\rangle\in
\mathcal{H}_{B}\;$. Action of the symmetric group occurs a decomposition of
the tensored space $\mathcal{H}^{\otimes n}$ ~\cite{Weyl},
\[
\mathcal{H}^{\otimes n}=\bigoplus_{\lambda}\mathcal{W}_{\lambda}%
,\;\mathcal{W}_{\lambda}:=\mathcal{U}_{\lambda}\otimes\mathcal{V}_{\lambda}.
\]
Here, $\mathcal{U}_{\lambda}$ and $\mathcal{V}_{\lambda}$ is an irreducible
space of the tensor representation of $\mathrm{SU}(d)$, and the representation
(\ref{sym}) of the symmetric group, respectively, and
\[
\lambda=(\lambda_{1},\ldots,\lambda_{d}),\quad\;\lambda_{i}\geq\lambda
_{i+1}\geq0,\,\sum_{i=1}^{d}\lambda_{i}=n
\]
is called \textit{Young index}, which $\mathcal{U}_{\lambda}$ and
$\mathcal{V}_{\lambda}$ uniquely corresponds to. To emphasize $\sum_{i=1}%
^{d}\lambda_{i}=n$, we use the notation " $\lambda\vdash n$".  We denote by
$\mathcal{U}_{\lambda,A}$, $\mathcal{V}_{\lambda,A}$, and $\mathcal{U}%
_{\lambda,B}$, $\mathcal{V}_{\lambda,B}$ the irreducible component of
$\mathcal{H}_{A}^{\otimes n}$ and $\mathcal{H}_{B}^{\otimes n}$ ,
respectively. Also, $\mathcal{W}_{\lambda,A}:=\mathcal{U}_{\lambda,A}%
\otimes\mathcal{V}_{\lambda,A}$, $\mathcal{W}_{\lambda,B}:=\mathcal{U}%
_{\lambda,B}\otimes\mathcal{V}_{\lambda,B}$.

In terms of \ this decomposition, $\left\vert \phi\right\rangle ^{\otimes n}$
can be written as%
\[
\left\vert \phi\right\rangle ^{\otimes n}=\bigoplus_{\lambda:\lambda\vdash
n}a_{\lambda}\left\vert \phi_{\lambda}\right\rangle \left\vert \Phi_{\lambda
}\right\rangle ,
\]

where $\left\vert \phi_{\lambda}\right\rangle \in\mathcal{U}_{\lambda
,A}\otimes\mathcal{U}_{\lambda,B}$, and $\left\vert \Phi_{\lambda
}\right\rangle \in\mathcal{V}_{\lambda,A}\otimes\mathcal{V}_{\lambda,B}$.
While $a_{\lambda}$ and $\left\vert \phi_{\lambda}\right\rangle $ are
dependent on $\left\vert \phi\right\rangle $, $\left\vert \Phi_{\lambda
}\right\rangle $ is a maximally entangled state which does not depend on
$\left\vert \phi\right\rangle $,%
\[
\left\vert \Phi_{\lambda}\right\rangle :=\frac{1}{\sqrt{d_{\lambda}}}%
\sum_{i=1}^{d_{\lambda}}\left\vert f_{i}\right\rangle \left\vert
f_{i}\right\rangle ,
\]
with $\left\{  \left\vert f_{i}\right\rangle \right\}  $'s being an
orthonormal complete basis of $\mathcal{V}_{\lambda}$, and $d_{\lambda}%
:=\dim\mathcal{V}_{\lambda}$.

Therefore, any symmetric pure state, being a superposition of $n$-tensored
pure states, can be written as%
\begin{equation}
\bigoplus_{\lambda:\lambda\vdash n}a_{\lambda}\left\vert \phi_{\lambda
}\right\rangle \left\vert \Phi_{\lambda}\right\rangle \,.\label{decomposition}%
\end{equation}

\section{Entanglement cost of symmetric states}

For Young indices $\lambda\vdash n$  and $l$ with $1\leq l\leq d_{\lambda}$ ,
let
\[
b_{\lambda l}^{n}=\frac{\mathrm{tr}\,\rho^{n}\mathcal{W}_{\lambda,A}%
\otimes\mathbf{1}_{B}}{d_{\lambda}}.
\]
Note $b_{\lambda l}^{n}$ does not vary with $l$.  Note also $b_{\lambda l}%
^{n}$ defines a probability distribution over $\left(  \lambda,k\right)  $.

\begin{lemma}
\label{lem:Ec-sym-lower}If the state is supported on the symmetric subspace of
$\left(  \mathcal{H}_{A}\otimes\mathcal{H}_{B}\right)  ^{\otimes n}$,
\begin{equation}
\mathrm{p-}\varlimsup_{n\rightarrow\infty}\frac{-1}{n}\log b_{\lambda l}%
^{n}\leq E_{c}\left(  \left\{  \rho^{n}\right\}  _{n=1}^{\infty}\right)
,\label{Ec-sym-lower}%
\end{equation}
where the underlying sequence of probability measure is $\left\{  b_{\lambda
k}^{n}\right\}  _{n=1}^{\infty}$\bigskip.
\end{lemma}

\begin{proof}
A composition of local projective measurement $\left\{  \mathcal{W}%
_{\lambda,A}\otimes\mathcal{W}_{\lambda,B}\right\}  _{\lambda}$ followed by
tracing out $\mathcal{U}_{A}\otimes\mathcal{U}_{B}$ sends symmetric state
$\rho^{n}$, a convex combination of a state in the form of
(\ref{decomposition}), to
\begin{equation}
\sigma^{n}:=\bigoplus_{\lambda:\lambda\vdash n}\sum_{l=1}^{d_{\lambda}%
}b_{\lambda l}^{n}\underset{}{\left\vert \Phi_{\lambda}\right\rangle
\left\langle \Phi_{\lambda}\right\vert },\label{random-unitary}%
\end{equation}
where $\left\vert \Phi_{\lambda}\right\rangle $ is a maximally entangled state
living in $\mathcal{V}_{\lambda,A}\otimes\mathcal{V}_{\lambda,B}$. Since this
operation is LOCC, dilution of $\sigma^{n}$ is easier than $\rho_{\phi}^{n,m}%
$, and
\begin{equation}
\mathrm{F}^{2^{nR}}\left(  \sigma^{n}\right)  \geq\mathrm{F}^{2^{nR}}\left(
\rho^{n}\right)  .\label{sigma-easier}%
\end{equation}
Let
\begin{align*}
\mathfrak{E}_{R}^{n} &  :=\left\{  \left(  \lambda,l\right)  ;\,b_{\lambda
l}^{n}\geq2^{-nR}\right\}  ,\\
\mathfrak{F}_{R}^{n} &  :=\left\{  \left(  \lambda,l\right)  ;\,d_{\lambda
}\leq2^{nR}\right\}  .
\end{align*}
Observe
\[
\mathrm{F}^{2^{nR}}\left(  \sigma^{n}\right)  =\sum_{\left(  \lambda,l\right)
\in\mathfrak{F}_{R^{n}}^{n}}b_{\lambda k}^{n}\,,
\]
where $R^{n}$ is decided by
\begin{equation}
R^{n}=\max\left\{  \,S\,;\sum_{\lambda\,:\,d_{\lambda}\leq2^{nS}}d_{\lambda
}\leq2^{nR}\,\right\}  .\label{def-sn}%
\end{equation}
Here, note that $R^{n}$ is a function of $R$, though we don't write it
explicitly. Since
\[
2^{nS}\leq\sum_{\lambda\,:\,d_{\lambda}\leq2^{nS}}d_{\lambda}\leq\left(
n+1\right)  ^{d}2^{nS},
\]
we have
\begin{equation}
R-\frac{d}{n}\log\left(  n+1\right)  \leq R^{n}\leq R.\label{Rn-R}%
\end{equation}
Since
\[
\mathfrak{F}_{R^{n}}^{n}\subset\mathfrak{F}_{R}^{n}\subset\left(
\mathfrak{F}_{R}^{n}\cap\overline{\mathfrak{E}_{R+\gamma}^{n}}\right)
\cup\mathfrak{E}_{R+\gamma}^{n}%
\]
holds,%
\[
\sum_{\left(  \lambda,l\right)  \in\mathfrak{F}_{R^{n}}^{n}}b_{\lambda l}%
^{n}\leq\sum_{\left(  \lambda,l\right)  \in\mathfrak{F}_{R}^{n}\cap
\overline{\mathfrak{E}_{R+\gamma}^{n}}}b_{\lambda l}^{n}+\sum_{\left(
\lambda,l\right)  \in\mathfrak{E}_{R+\gamma}^{n}}b_{\lambda l}^{n}.
\]
We show the first term of the right hand side is negligible for any $\gamma
>0$:
\begin{align*}
\sum_{\left(  \lambda,l\right)  \in\mathfrak{F}_{R}^{n}\cap\overline
{\mathfrak{E}_{R}^{n}}}b_{\lambda k}^{n} &  \leq2^{-n\left(  R+\gamma\right)
}\left\vert \left\{  \left(  \lambda,l\right)  \,;d_{\lambda}\leq2^{nR^{n}%
}\,\,\right\}  \right\vert \\
&  =\,2^{-n\left(  R+\gamma\right)  }\sum_{\lambda:d_{\lambda}\leq2^{nR}%
}d_{\lambda}\\
&  \leq2^{-n\left(  R+\gamma\right)  }\cdot2^{nR}\cdot\left(  n+1\right)
^{d}\\
&  \leq\left(  n+1\right)  ^{d}2^{-n\gamma}.
\end{align*}
Therefore,
\begin{equation}
\mathrm{F}^{2^{nR}}\left(  \sigma^{n}\right)  \leq\sum_{\left(  \lambda
,l\right)  \in\mathfrak{E}_{R+\gamma}^{n}}b_{\lambda l}^{n}+\left(
n+1\right)  ^{d}2^{-n\gamma}.\label{opt-fidel-upper-sigma}%
\end{equation}
which, combined with (\ref{sigma-easier}),  implies
\begin{align*}
E_{c}\left(  \left\{  \rho^{n}\right\}  _{n=1}^{\infty}\right)   &
=\inf\left\{  R\,\,;\lim_{n\rightarrow\infty}\mathrm{F}^{2^{nR}}\left(
\rho^{n}\right)  =1\right\}  \\
&  \geq\inf\left\{  R\,\,;\lim_{n\rightarrow\infty}\mathrm{F}^{2^{nR}}\left(
\sigma^{n}\right)  =1\right\}  \\
&  \geq\inf\left\{  R\,\,;\lim_{n\rightarrow\infty}\,\,\sum_{\left(
\lambda,l\right)  \in\mathfrak{E}_{R+\gamma}^{n}}b_{\lambda l}^{n}+\left(
n+1\right)  ^{d}2^{-n\gamma}=1\right\}  \\
&  =\inf\left\{  R\,\,;\lim_{n\rightarrow\infty}\,\sum_{\left(  \lambda
,l\right)  \in\mathfrak{E}_{R+\gamma}^{n}}b_{\lambda l}^{n}=1\right\}  \\
&  =\mathrm{p-}\varlimsup_{n\rightarrow\infty}\frac{-1}{n}\log b_{\lambda
l}^{n}-\gamma.
\end{align*}
Since this holds for any $\gamma>0$, our assertion is proved.
\end{proof}

\bigskip

Below, we present a dilation protocol achieving the left hand side of
(\ref{Ec-sym-lower}). First, Bob fabricates the state locally, and applies the
binary projective measurement
\[
\left\{  \sum_{\lambda:d_{\lambda}\leq2^{nR}}\mathcal{W}_{\lambda
,B},\,\mathbf{1}-\sum_{\lambda:d_{\lambda}\leq2^{nR}}\mathcal{W}_{\lambda
,B}\right\}  ,
\]
where $R=\mathrm{p-}\varlimsup_{n\rightarrow\infty}\frac{-1}{n}\log b_{\lambda
l}^{n}+\gamma$ ($\gamma>0$). If the event corresponding to $\sum
_{\lambda:d_{\lambda}\leq2^{nR}}\mathcal{W}_{\lambda,B}$ is observed, he
teleports the part which should belong to Alice.

This procedure consumes the following amount of entanglement:%

\begin{align*}
&  \log\sum_{d_{\lambda}\leq2^{nR}}d_{\lambda}\dim\mathcal{U}_{\lambda}\\
&  \leq nR+d\log\left(  n+1\right)  +d^{2}\log n,
\end{align*}
(see (\ref{dim-zero-rate}) ). Dividing both ends by $n$ $\ $and taking
$\varlimsup_{n\rightarrow\infty}$, the left hand side becomes $R$, which can
be arbitrarily close to $\mathrm{p-}\varlimsup_{n\rightarrow\infty}\frac
{-1}{n}\log b_{\lambda l}^{n}$. \ The success fidelity of this protocol is
\[
\sum_{\left(  \lambda,l\right)  \in\mathfrak{F}_{R}^{n}}b_{\lambda l}^{n}\,.
\]
Since $1\geq d_{\lambda}b_{\lambda l}^{n}$ or $d_{\lambda}\leq\left(
b_{\lambda l}^{n}\right)  ^{-1}$, we have $\mathfrak{F}_{R}^{n}\supset
\mathfrak{E}_{R}^{n}$ \ and
\begin{equation}
\sum_{\left(  \lambda,l\right)  \in\mathfrak{F}_{R}^{n}}b_{\lambda l}%
^{n}\,\geq\sum_{\left(  \lambda,l\right)  \in\mathfrak{E}_{R}^{n}}b_{\lambda
l}^{n},\label{sum-dim>sum-prob}%
\end{equation}
which tends to $1$ since $R>\mathrm{p-}\varlimsup_{n\rightarrow\infty}%
\frac{-1}{n}\log b_{\lambda l}^{n}$. Therefore, combined with lemma\thinspace
\ref{lem:Ec-sym-lower}, we have proved:

\begin{theorem}
\label{th:Ec-sym} If $\rho^{n}$ is a symmetric state,
\[
E_{c}\left(  \left\{  \rho^{n}\right\}  _{n=1}^{\infty}\right)  =\mathrm{p-}%
\varlimsup_{n\rightarrow\infty}\frac{-1}{n}\log b_{\lambda l}^{n},
\]
where the underlying sequence of probability measure is $\left\{  b_{\lambda
l}^{n}\right\}  _{n=1}^{\infty}$. $E_{c}\left(  \left\{  \rho^{n}\right\}
_{n=1}^{\infty}\right)  $ can be achieved by creating state locally and
teleporting it.
\end{theorem}

\bigskip\ 

Below, we derive another expression of $E_{c}\left(  \left\{  \rho
^{n}\right\}  _{n=1}^{\infty}\right)  $. Let $\left\{  c_{\lambda\,k\,l}%
^{n}\right\}  $ be spectrum of the reduced density matrix $\mathrm{tr}%
_{\mathcal{H}_{A}}\rho^{n}$.
\[
\mathrm{tr}_{\mathcal{H}_{A}}\rho^{n}=\sum_{\lambda,\,k,\,l}c_{\lambda
\,k\,l}^{n}\left\vert \lambda\,k\,l\right\rangle \left\langle \lambda
\,k\,l\,\right\vert ,
\]
where $\left\vert \lambda\,k\,l\right\rangle \in\mathcal{W}_{\lambda}$, and
the indices $k$ and $l$ corresponds to the freedom of $\mathcal{U}_{\lambda}$
and $\mathcal{V}_{\lambda}$, respectively. Note that the Schmidt coefficient
$c_{\lambda kl}^{n}$ does not depend on $l$, and that
\[
\sum_{k=1}^{\dim\mathcal{U}_{\lambda}}c_{\lambda kl}^{n}=b_{\lambda l}^{n}\,.
\]

\begin{theorem}
\label{th:Ec-sym-2}%
\[
\mathrm{p-}\varlimsup_{n\rightarrow\infty}\frac{-1}{n}\log c_{\lambda kl}%
^{n}=E_{c}\left(  \left\{  \rho^{n}\right\}  _{n=1}^{\infty}\right)  ,
\]
where the probabilistic limsup is with respect to the sequence of probability
measure $\left\{  c_{\lambda kl}^{n}\right\}  _{n=1}^{\infty}$.
\end{theorem}

\begin{proof}
Due to the definition of probabilistic limsup,%
\begin{align}
E_{c}\left(  \left\{  \rho^{n}\right\}  _{n=1}^{\infty}\right)   &
=\mathrm{p-}\varlimsup_{n\rightarrow\infty}\frac{-1}{n}\log\sum_{k=1}%
^{\dim\mathcal{U}_{\lambda}}c_{\lambda kl}^{n}\text{ \ (w.r.t. }\left\{
b_{\lambda l}^{n}\right\}  \text{)}\label{Ec-upper}\\
&  =\mathrm{p-}\varlimsup_{n\rightarrow\infty}\frac{-1}{n}\log\sum_{k^{\prime
}=1}^{\dim\mathcal{U}_{\lambda}}c_{\lambda k^{\prime}l}^{n}\text{ \ (w.r.t.
}\left\{  c_{\lambda kl}^{n}\right\}  \text{)}\nonumber\\
&  \leq\mathrm{p-}\varlimsup_{n\rightarrow\infty}\frac{-1}{n}\log c_{\lambda
kl}^{n}\text{ \ (w.r.t. }\left\{  c_{\lambda kl}^{n}\right\}  \text{)}%
.\nonumber
\end{align}
On the other hand,
\begin{align*}
&  \mathrm{p-}\varlimsup_{n\rightarrow\infty}\frac{-1}{n}\log c_{\lambda
kl}^{n}\,\ \text{(w.r.t. }\left\{  c_{\lambda kl}^{n}\right\}  \text{)}\,\\
&  \leq\mathrm{p-}\varlimsup_{n\rightarrow\infty}\frac{-1}{n}\log b_{\lambda
l}^{n}\,\text{(w.r.t. }\left\{  c_{\lambda kl}^{n}\right\}  \text{)}\\
&  +\,\mathrm{p-}\varlimsup_{n\rightarrow\infty}\frac{-1}{n}\log\left(
c_{\lambda kl}^{n}\,/b_{\lambda l}^{n}\right)  \,\,\text{(w.r.t. }\left\{
c_{\lambda kl}^{n}\right\}  \text{)}.
\end{align*}
Observe that $\left\{  c_{\lambda kl}^{n}\,/b_{\lambda l}^{n}\right\}
_{k=1}^{\dim\mathcal{U}_{\lambda}}$ defines a probability distribution over
$k\,$ ($1\leq k\leq\dim\mathcal{U}_{\lambda}$). Letting $R$ and $\gamma$ be an
arbitrary positive real number, we have%
\begin{align*}
& \sum_{\lambda,k,l\,:\,\left(  c_{\lambda kl}^{n}\,/b_{\lambda l}^{n}\right)
\geq2^{-n\left(  R+\gamma\right)  }}c_{\lambda kl}^{n}\\
& =\sum_{\lambda,l}\left[  b_{\lambda l}^{n}\sum_{k\,:\,\left(  c_{\lambda
kl}^{n}\,/b_{\lambda l}^{n}\right)  \geq2^{-n\left(  R+\gamma\right)  }}%
\frac{c_{\lambda kl}^{n}\,}{b_{\lambda l}^{n}}\right]  \\
& \geq\sum_{\lambda,l}\left[  b_{\lambda l}^{n}\sum_{k\,:\,k\,\leq
2^{nR},\,\left(  c_{\lambda kl}^{n}\,/b_{\lambda l}^{n}\right)  \geq
2^{-n\left(  R+\gamma\right)  }}\frac{c_{\lambda kl}^{n}\,}{b_{\lambda l}^{n}%
}\right]  \\
& =\sum_{\lambda,l}\left[  b_{\lambda l}^{n}\left\{  \sum_{k\,:\,k\,\leq
2^{nR}}\frac{c_{\lambda kl}^{n}\,}{b_{\lambda l}^{n}}-\sum_{k\,:\,k\,\leq
2^{nR},\,\left(  c_{\lambda kl}^{n}\,/b_{\lambda l}^{n}\right)  <2^{-n\left(
R+\gamma\right)  }}\frac{c_{\lambda kl}^{n}\,}{b_{\lambda l}^{n}}\right\}
\right]  \\
& \geq\sum_{\lambda,l}\left[  b_{\lambda l}^{n}\left\{  \sum_{k\,:\,k\,\leq
2^{nR}}\frac{c_{\lambda kl}^{n}\,}{b_{\lambda l}^{n}}-\sum_{k\,:\,k\,\leq
2^{nR}}2^{-n\left(  R+\gamma\right)  }\right\}  \right]  \\
& =\sum_{\lambda,l}b_{\lambda l}^{n}\sum_{k\,:\,k\,\leq2^{nR}}\frac{c_{\lambda
kl}^{n}\,}{b_{\lambda l}^{n}}-2^{-n\gamma}.
\end{align*}
Since $2^{nR}\geq\dim\mathcal{U}_{\lambda}$ holds for any $R>0$ with large
$n$, \ the last end of the inequality converges to $1$. Hence, the left most
end converges to $1$, also. This means
\[
\mathrm{p-}\varlimsup_{n\rightarrow\infty}\frac{-1}{n}\log\left(  c_{\lambda
kl}^{n}\,/b_{\lambda l}^{n}\right)  \,\,\text{(w.r.t. }\left\{  c_{\lambda
kl}^{n}\right\}  \text{)}\leq R+\gamma.
\]
Letting $R\rightarrow0$ and $\gamma\rightarrow0$, we obtain
\begin{equation}
\mathrm{p-}\varlimsup_{n\rightarrow\infty}\frac{-1}{n}\log\left(  c_{\lambda
kl}^{n}\,/b_{\lambda l}^{n}\right)  \,\,\text{(w.r.t. }\left\{  c_{\lambda
kl}^{n}\right\}  \text{)}=0\label{p-limsup-log(c/b)=0}%
\end{equation}
and%
\begin{align*}
&  \mathrm{p-}\varlimsup_{n\rightarrow\infty}\frac{-1}{n}\log c_{\lambda
kl}^{n}\leq\mathrm{p-}\varlimsup_{n\rightarrow\infty}\frac{-1}{n}\log
b_{\lambda l}^{n}\,\\
&  =E_{c}\left(  \left\{  \rho^{n}\right\}  _{n=0}^{\infty}\right)  .
\end{align*}
Combining this with (\ref{Ec-upper}), we have the assertion.
\end{proof}

\section{Distillable entanglement of symmetric states}

\begin{lemma}
\label{Ed-lower}If $\rho^{n}$ is a symmetric state,
\[
E_{d}\left(  \left\{  \rho^{n}\right\}  _{n=0}^{\infty}\right)  \geq
\mathrm{p-}\varliminf_{n\rightarrow\infty}\frac{-1}{n}\log b_{\lambda l}%
^{n}\,\,,
\]
where the probabilistic limsup is with respect to the sequence of probability
measure $\left\{  b_{\lambda l}^{n}\right\}  _{n=1}^{\infty}$. Especially, the
right hand side can be achieved without knowing $\rho^{n}$, expect the fact
that it is a symmetric state.
\end{lemma}

\begin{proof}
Alice and Bob applies $\left\{  \mathcal{W}_{\lambda,A}\right\}  _{\lambda}$
and $\left\{  \mathcal{W}_{\lambda,B}\right\}  _{\lambda}$ independently, and
trace out $\mathcal{U}_{\lambda,A}$ and $\mathcal{U}_{\lambda,B}$,
respectively. Then, they obtain $\frac{1}{n}\log d_{\lambda}$ ebits of Bell
pairs with the probability $d_{\lambda}b_{\lambda l}^{n}\,$. They also obtain
classical information about $\lambda$, so they exactly know the shared
entangled state. Obviously, this protocol can be implemented without knowing
the input. Obviously, the yield of the protocol is
\[
\mathrm{p-}\varliminf_{n\rightarrow\infty}\frac{1}{n}\log d_{\lambda}=\sup
_{R}\left\{  R\,\,;\lim_{n\rightarrow\infty}\,\sum_{\left(  \lambda,l\right)
\in\mathfrak{F}_{R}^{n}}b_{\lambda l}^{n}=0\right\}  .
\]
If the sum over $\mathfrak{F}_{R}^{n}$ can be replaced by $\mathfrak{E}%
_{R}^{n}$, we are done. Since
\[
\mathfrak{F}_{R}^{n}\subset\left(  \mathfrak{F}_{R}^{n}\cap\overline
{\mathfrak{E}_{R+\gamma}^{n}}\right)  \cup\mathfrak{E}_{R+\gamma}^{n}%
\]
holds,
\[
\sum_{\left(  \lambda,l\right)  \in\mathfrak{F}_{R}^{n}}b_{\lambda l}^{n}%
\leq\sum_{\left(  \lambda,l\right)  \in\mathfrak{F}_{R}^{n}\cap\overline
{\mathfrak{E}_{R+\gamma}^{n}}}b_{\lambda l}^{n}+\sum_{\left(  \lambda
,l\right)  \in\mathfrak{E}_{R+\gamma}^{n}}b_{\lambda l}^{n}.
\]
We show the first term of the right hand side is negligible for any $\gamma
>0$:
\begin{align*}
\sum_{\left(  \lambda,l\right)  \in\mathfrak{F}_{R}^{n}\cap\overline
{\mathfrak{E}_{R}^{n}}}b_{\lambda l}^{n} &  \leq2^{-n\left(  R+\gamma\right)
}\left\vert \left\{  \left(  \lambda,l\right)  \,;d_{\lambda}\leq
2^{nR}\,\,\right\}  \right\vert \\
&  =\,2^{-n\left(  R+\gamma\right)  }\sum_{\lambda:d_{\lambda}\leq2^{nR}%
}d_{\lambda}\\
&  \leq2^{-n\left(  R+\gamma\right)  }\cdot2^{nR}\cdot\left(  n+1\right)
^{d}\\
&  =\left(  n+1\right)  ^{d}2^{-n\gamma}.
\end{align*}
Therefore,
\begin{align*}
E_{d}\left(  \left\{  \rho^{n}\right\}  _{n=1}^{\infty}\right)   &  \geq
\sup_{R}\left\{  R\,\,;\lim_{n\rightarrow\infty}\,\left(  \sum_{\left(
\lambda,l\right)  \in\mathfrak{E}_{R+\gamma}^{n}}b_{\lambda l}^{n}+\left(
n+1\right)  ^{d}2^{-n\gamma}\right)  =0\right\}  \\
&  =\sup_{R}\left\{  R\,\,;\,\lim_{n\rightarrow\infty}\sum_{\left(
\lambda,l\right)  \in\mathfrak{E}_{R+\gamma}^{n}}b_{\lambda l}^{n}\leq
\epsilon\right\}  \\
&  =\mathrm{p-}\varliminf_{n\rightarrow\infty}\frac{-1}{n}\log b_{\lambda
l}^{n}-\gamma.
\end{align*}
Since this holds for all $\gamma$, the lemma is proven.
\end{proof}

\bigskip

\begin{lemma}
If Alice's view of $\left\vert \psi^{n}\right\rangle $ is the same as
$\rho^{n}$, i.e.,
\[
\mathrm{tr}\,_{\mathcal{H}_{B}}\left\vert \psi^{n}\right\rangle \left\langle
\psi^{n}\right\vert =\rho^{n},
\]%
\[
E_{d}\left(  \left\{  \rho^{n}\right\}  _{n=0}^{\infty}\right)  \leq
E_{d}\left(  \left\{  \left\vert \psi^{n}\right\rangle \right\}
_{n=0}^{\infty}\right)
\]

\end{lemma}

\begin{proof}
We prove that $\rho^{n}$ can be made from $\left\vert \psi^{n}\right\rangle $
by a local operation. Let $\left\vert \psi^{\prime}\right\rangle \in\left(
\mathcal{H}_{A}\otimes\mathcal{H}_{B}\right)  ^{\otimes n}\otimes\mathcal{K}$
be a purification of $\rho^{n}$. Since Alice's view of $\left\vert
\psi^{\prime}\right\rangle $ and $\left\vert \psi^{n}\right\rangle $ are the
same, $\left\vert \psi^{\prime}\right\rangle $ is mapped to $\left\vert
\psi^{n}\right\rangle $ by a local isometry acting on $\mathcal{H}%
_{B}^{\otimes n}\otimes$ $\mathcal{K}$.
\end{proof}

\begin{theorem}
\label{th:Ed-sym}If $\rho^{n}$ is supported on the symmetric subspace of
$\left(  \mathcal{H}_{A}\otimes\mathcal{H}_{B}\right)  ^{\otimes n}$,
\begin{align*}
E_{d}\left(  \left\{  \rho^{n}\right\}  _{n=0}^{\infty}\right)    &
=\mathrm{p-}\varliminf_{n\rightarrow\infty}\frac{-1}{n}\log b_{\lambda l}%
^{n}\\
& =\,\,\mathrm{p-}\varliminf_{n\rightarrow\infty}\frac{-1}{n}\log c_{\lambda
kl}^{n}.
\end{align*}

\end{theorem}

A remarkable point is that the optimal rate can be achieved without knowing
$\rho^{n}$, as is indecated in lemma \ref{Ed-lower}. This is a natural
generalization of universal entanglement concentration in
\cite{MatsumotoHayashi}.

\begin{proof}
Due to \cite{Hayashi:2006} and the above lemma,
\[
E_{d}\left(  \left\{  \rho^{n}\right\}  _{n=0}^{\infty}\right)  \leq
\mathrm{p-}\varliminf_{n\rightarrow\infty}\frac{-1}{n}\log c_{\lambda kl}^{n}.
\]
Due to lemma\thinspace\ref{Ed-lower},
\[
\mathrm{p-}\varliminf_{n\rightarrow\infty}\frac{-1}{n}\log b_{\lambda l}%
^{n}\leq\mathrm{p-}\varliminf_{n\rightarrow\infty}\frac{-1}{n}\log c_{\lambda
kl}^{n}.
\]
Hence, our task is only to show the opposite inequality:%
\begin{align*}
& \mathrm{p-}\varliminf_{n\rightarrow\infty}\frac{-1}{n}\log b_{\lambda l}%
^{n}\\
& =\mathrm{p-}\varliminf_{n\rightarrow\infty}\frac{-1}{n}\left[  \log
c_{\lambda kl}^{n}-\log\left(  c_{\lambda kl}^{n}\,/b_{\lambda l}^{n}\right)
\right]  \\
& \geq\mathrm{p-}\varliminf_{n\rightarrow\infty}\frac{-1}{n}\log c_{\lambda
kl}^{n}+\mathrm{p-}\varliminf_{n\rightarrow\infty}\frac{1}{n}\log\left(
c_{\lambda kl}^{n}\,/b_{\lambda l}^{n}\right)  \\
& =\mathrm{p-}\varliminf_{n\rightarrow\infty}\frac{-1}{n}\log c_{\lambda
kl}^{n}-\mathrm{p-}\varlimsup_{n\rightarrow\infty}\frac{-1}{n}\log\left(
c_{\lambda kl}^{n}\,/b_{\lambda l}^{n}\right)  \\
& =\mathrm{p-}\varliminf_{n\rightarrow\infty}\frac{-1}{n}\log c_{\lambda
kl}^{n},
\end{align*}
where the last equality is due to (\ref{p-limsup-log(c/b)=0}).
\end{proof}

\section{Strong converse}

The strong converse property for entanglement dilution is defined as follows:
If $\mathrm{F}^{2^{nR}}\left(  \rho^{n}\right)  \rightarrow0$ occurs for all
$R<E_{c}\left(  \left\{  \rho^{n}\right\}  _{n=0}^{\infty}\right)  $,  we say
that $\left\{  \rho^{n}\right\}  _{n=0}^{\infty}$ has strong converse property
for dilution

Similarly, the strong converse property for entanglement distillation is
defined as follows. Denote by $\mathrm{F}_{d}^{D}\left(  \rho\right)  $ the
optimal fidelity of making the maximally entangled state with Schmidt rank $D$
from $\rho$ by LOCC.  $\left\{  \rho^{n}\right\}  _{n=0}^{\infty}$ is said to
have strong converse property for distillation if $\mathrm{F}_{d}^{2^{nR}%
}\left(  \rho^{n}\right)  \rightarrow0$ occurs for all $R>E_{d}\left(
\left\{  \rho^{n}\right\}  _{n=0}^{\infty}\right)  $.   

\begin{theorem}
Suppose $\rho^{n}$ is a symmetric state.\quad Then, the following three
conditions are equivalent. \ (i) Entanglement dilution of $\left\{  \rho
^{n}\right\}  _{n=0}^{\infty}$ has strong converse property. (ii) Entanglement
distillation from $\left\{  \rho^{n}\right\}  _{n=0}^{\infty}$ has strong
converse property. (iii) $E_{c}\left(  \left\{  \rho^{n}\right\}
_{n=0}^{\infty}\right)  =E_{d}\left(  \left\{  \rho^{n}\right\}
_{n=0}^{\infty}\right)  $.
\end{theorem}

\begin{proof}
First we prove (i)$\Leftarrow$(iii). Combination of (\ref{sigma-easier}) and
(\ref{opt-fidel-upper-sigma}) yields%
\[
\mathrm{F}^{2^{nR}}\left(  \rho^{n}\right)  \leq\left(  n+1\right)
^{d}2^{-n\gamma}+\sum_{\left(  \lambda,l\right)  \in\mathfrak{E}_{R+\gamma
}^{n}}b_{\lambda l}^{n}.
\]
Hence, if
\[
R+\gamma<\mathrm{p-}\varliminf_{n\rightarrow\infty}\frac{-1}{n}\log b_{\lambda
l}^{n}=E_{d}\left(  \left\{  \rho^{n}\right\}  _{n=0}^{\infty}\right)  ,
\]
the last end asymptotically vanishes. Since $\gamma>0$ is arbitrary, we obtain
(i)$\Leftarrow$(iii). On the other hand, if (i) holds, the entanglement
dilution protocol mentioned right before the theorem\thinspace\ref{th:Ec-sym}
also can achieve only asymptotically vanishing fidelity with  $R<E_{c}\left(
\left\{  \rho^{n}\right\}  _{n=0}^{\infty}\right)  $ :
\[
\lim_{n\rightarrow\infty}\sum_{\left(  \lambda,l\right)  \in\mathfrak{F}%
_{R^{n}}^{n}}b_{\lambda l}^{n}=0,
\]
where $R^{n}$ is defined by (\ref{def-sn}). Due to (\ref{sum-dim>sum-prob}),
this implies
\[
\lim_{n\rightarrow\infty}\sum_{\left(  \lambda,l\right)  \in\mathfrak{E}%
_{R^{n}}^{n}}b_{\lambda l}^{n}=0.
\]
Since $R^{n}\rightarrow R$ as $n\rightarrow\infty$ due to (\ref{Rn-R}), this
implies
\[
R\leq\mathrm{p-}\varliminf_{n\rightarrow\infty}\frac{-1}{n}\log b_{\lambda
l}^{n}=E_{d}\left(  \left\{  \rho^{n}\right\}  _{n=0}^{\infty}\right)  .
\]
Therefore, we have (i)$\Rightarrow$(iii). \ Next, we suppose that (ii) holds.
Then, with $R>E_{d}\left(  \left\{  \rho^{n}\right\}  _{n=0}^{\infty}\right)
$, the protocol in the proof of lemma\thinspace\ref{Ed-lower} can achieve only
asymptotically vanishing fidelity :%
\[
\lim_{n\rightarrow\infty}\sum_{\left(  \lambda,l\right)  \notin\mathfrak{F}%
_{R^{n}}^{n}}b_{\lambda l}^{n}=0.
\]
Observe
\begin{align*}
1-\sum_{\left(  \lambda,l\right)  \notin\mathfrak{F}_{R^{n}}^{n}}b_{\lambda
l}^{n}  & =\sum_{\left(  \lambda,l\right)  \in\mathfrak{F}_{R^{n}}^{n}%
}b_{\lambda l}^{n}\\
& =\mathrm{F}^{2^{nR^{n}}}\left(  \sigma^{n}\right)  \\
& \leq\mathrm{F}^{2^{nR^{n}}}\left(  \rho^{n}\right)  \\
& \leq\mathrm{F}^{2^{nR}}\left(  \rho^{n}\right)  \\
& \leq\left(  n+1\right)  ^{d}2^{-n\gamma}+\sum_{\left(  \lambda,l\right)
\in\mathfrak{E}_{R+\gamma}^{n}}b_{\lambda l}^{n},
\end{align*}
where the inequality in the third, fourth, and the last line is due to
(\ref{sigma-easier}), (\ref{Rn-R}), and (\ref{opt-fidel-upper-sigma}),
respectively.  Therefore, we have
\[
\lim_{n\rightarrow\infty}\sum_{\left(  \lambda,l\right)  \in\mathfrak{E}%
_{R+\gamma}^{n}}b_{\lambda l}^{n}=1
\]
for any $\gamma>0$, or equivalently,
\[
R\geq\mathrm{p-}\varlimsup_{n\rightarrow\infty}\frac{-1}{n}\log b_{\lambda
l}^{n}=E_{c}\left(  \left\{  \rho^{n}\right\}  _{n=0}^{\infty}\right)  .
\]
Therefore, we have (ii)$\Rightarrow$(iii). Finally, we show (ii)$\Leftarrow
$(iii). Let $\left\vert \psi^{n}\right\rangle $ be a purification of $\rho
^{n}$, with all the ancilla at Bob's hand. Obviously,
\[
\mathrm{F}_{d}^{2^{nR}}\left(  \left\vert \psi^{n}\right\rangle \right)
\geq\mathrm{F}_{d}^{2^{nR}}\left(  \rho^{n}\right)  .
\]
\ Suppose $R>\mathrm{p-}\varlimsup_{n\rightarrow\infty}\frac{-1}{n}\log
c_{\lambda kl}^{n}$. Then, \cite{Hayashi:2006} had shown that $\mathrm{F}%
_{d}^{2^{nR}}\left(  \left\vert \psi^{n}\right\rangle \right)  \rightarrow0$.
\[
\mathrm{p-}\varlimsup_{n\rightarrow\infty}\frac{-1}{n}\log c_{\lambda kl}%
^{n}=E_{c}\left(  \left\{  \rho^{n}\right\}  _{n=0}^{\infty}\right)
=E_{d}\left(  \left\{  \rho^{n}\right\}  _{n=0}^{\infty}\right)
\]
holds by assumption, this implies strong converse for distillation.
\end{proof}

\section{Output of an optimal cloning machine (1)}

In this and next section, we study the ouput states of  cloning machines. They
are, if optimally desined for pure input states, mixed symmetric states.

In this section, we suppose that the Schmidt basis of the given pure state,
except for its phases, are known i.e.,
\[
\left\vert \phi\right\rangle =\sum_{i=1}^{d}\sqrt{p_{i}}e^{\sqrt{-1}\theta
_{i}}\left\vert i\right\rangle \left\vert i\right\rangle ,
\]
where $\boldsymbol{p}=\left(  p_{1},\cdots,p_{d}\right)  $ and $\theta_{i}$
($i=1,\cdots,d$) are unknown. The final state of Its optimal $n$ to $m$
\ \ cloning machine is%

\[
\sum_{\boldsymbol{n},\boldsymbol{m}}\alpha_{\boldsymbol{m,n}}\left\vert
\boldsymbol{m}\right\rangle \otimes\left\vert R_{\boldsymbol{m}-\boldsymbol{n}%
}\right\rangle ,
\]
where \
\[
\left\vert \boldsymbol{m}\right\rangle =\sqrt{\frac{\prod_{k=1}^{d}m_{k}!}%
{m!}}\sum_{\#\{\,\kappa\,;\,i_{\kappa}=k\}=m_{k}}\bigotimes_{\kappa=1}%
^{m}\left\vert i_{\kappa}\right\rangle \left\vert i_{\kappa}\right\rangle \,,
\]%
\[
\alpha_{\boldsymbol{m,n}}=\sqrt{\frac{\left(  m-n\right)  !\left(
n+d-1\right)  !}{\left(  m+d-1\right)  !}}\prod_{k=1}^{d}\sqrt{\frac{m_{k}%
!}{n_{k}!\,\left(  m_{k}-n_{k}\right)  !}}\sqrt{\frac{n!}{\prod_{k=1}^{d}%
n_{k}!}p_{k}^{n_{k}}}e^{\sqrt{-1}\theta_{k}}%
\]
and $\left\{  \left\vert R_{\boldsymbol{j}}\right\rangle \right\}  $ is an
orthonormal basis of the internal state of the optimal cloning
machine\thinspace\cite{FanMatsumoto}. Tracing out the internal state of the
machine, we obtain the output state, which is denoted by $\rho_{1}^{n,m}$.
Below, we denote by $H\left(  \boldsymbol{p}\right)  $ the Shannon entropy of
the probability distribution $\boldsymbol{p}$.

\begin{theorem}
\label{th:EcEdclone}%
\begin{equation}
E_{c}\left(  \left\{  \rho_{1}^{m/r,m}\right\}  _{m=1}^{\infty}\right)
=E_{d}\left(  \left\{  \rho_{1}^{m/r,m}\right\}  _{m=1}^{\infty}\right)
=H\left(  \boldsymbol{p}\right)  .\label{EcEdclone}%
\end{equation}

\end{theorem}

An important consequence of this is that the strong converse holds for
$\left\{  \rho_{1}^{m/r,m}\right\}  _{m=1}^{\infty}$.

Also, the real $m$ copies and optimal clone are the same in entanglement
quantities. (Recall all the reasonable entanglement measures lies between
$E_{d}$ and $E_{c}$.) \ This is rather surprising since $\mathrm{F}\left(
\rho_{1}^{m/r,m},\left\vert \phi\right\rangle ^{\otimes m}\right)  \approx
r^{d}$, and these two states are not so close.

However, with closer look, entanglement of $\rho_{1}^{m/r,m}$ and $\left\vert
\phi\right\rangle ^{\otimes m}$ are somewhat different. More concretely, they
differ in error exponent of entanglement dilution. Below, $h(x):=-x\log
x-(1-x)\log(1-x)$.

\begin{theorem}
\label{th:exponent-dilution-clone}If $R>H\left(  \boldsymbol{p}\right)  $,
\quad%
\begin{align}
& \lim_{m\rightarrow\infty}\frac{-1}{m}\log\left\{  1-\mathrm{F}^{2^{mR}%
}\left(  \rho_{1}^{m/r,m}\right)  \right\}  \nonumber\\
& =\min_{\boldsymbol{q}:\,H\left(  \boldsymbol{q}\right)  \geq R}%
\min_{\boldsymbol{q}^{\prime}:q_{i}^{\prime}\leq rq_{i}}\left\{  h\left(
\frac{1}{r}\right)  -\sum_{i=1}^{d}q_{i}\,h\left(  \frac{q_{i}^{\prime}%
}{rq_{i}}\right)  +\frac{1}{r}D\left(  \boldsymbol{q}^{\prime}%
||\,\boldsymbol{p}\right)  \right\}  .\label{exponent-dilution-clone}%
\end{align}
Equivalently, if \ $\lim_{m\rightarrow\infty}\frac{-1}{m}\log\left\{
1-\mathrm{F}^{2^{mR}}\left(  \rho_{1}^{m/r,m}\right)  \right\}  \geq\eta$,
\ we at least need following ebits of maximally entangled state:
\[
\max\left\{  H\left(  \boldsymbol{q}\right)  \,;\min_{\boldsymbol{q}^{\prime
}:q_{i}^{\prime}\leq rq_{i}}\left\{  h\left(  \frac{1}{r}\right)  -\sum
_{i=1}^{d}q_{i}\,h\left(  \frac{q_{i}^{\prime}}{rq_{i}}\right)  +\frac{1}%
{r}D\left(  \boldsymbol{q}^{\prime}||\,\boldsymbol{p}\right)  \right\}
\geq\eta\right\}  .
\]

\end{theorem}

Observe that (\ref{exponent-dilution-clone}) is smaller than or equal to
\[
\frac{1}{r}\min_{\boldsymbol{q}:\,H\left(  \boldsymbol{q}\right)  \geq
R}D\left(  \boldsymbol{q}||\,\boldsymbol{p}\right)  .
\]
It is known that the exponent for $\left\vert \phi\right\rangle ^{\otimes m}%
$,
\[
\lim_{m\rightarrow\infty}\frac{-1}{m}\log\left\{  1-\mathrm{F}^{2^{mR}}\left(
\left\vert \phi\right\rangle ^{\otimes m}\right)  \right\}  =\min
_{\boldsymbol{q}:\,H\left(  \boldsymbol{q}\right)  \geq R}D\left(
\boldsymbol{q}||\,\boldsymbol{p}\right)
\]
(see \cite{Hayashi:Springer}). Therefore, (\ref{exponent-dilution-clone}) is
smaller than or equal to  the exponent for $\left\vert \phi\right\rangle
^{\otimes\left(  m/r\right)  }$, or the input of the cloning machine.

\subsection{Proof of theorem\thinspace\ref{th:EcEdclone}}

The eigenvectors of the reduced density matrix $\mathrm{tr}\,_{\mathcal{H}%
_{B}^{\otimes m}}\rho_{1}^{n,m}$ are%
\[
\left\vert i_{1}\cdots i_{m}\right\rangle :=\bigotimes_{\kappa=1}%
^{m}\left\vert i_{\kappa}\right\rangle
\]
with corresponding eigenvalues%
\[
\frac{\prod_{k=1}^{d}m_{k}!}{m!}\sum_{\boldsymbol{n}}\left\vert \alpha
_{\boldsymbol{m,n}}\right\vert ^{2},
\]
where $\#\{\,\kappa\,;\,i_{\kappa}=k\}=m_{k}$. Each eigenvalue has
$m!/\prod_{k=1}^{d}m_{k}!$ folds degeneracy.

Due to theorem\thinspace\ref{th:Ec-sym-2}, $R\geq E_{c}\left(  \left\{
\rho_{1}^{n,rn}\right\}  _{n=1}^{\infty}\right)  $ holds if and only if%
\[
\lim_{n\rightarrow\infty}\sum_{\boldsymbol{m}\notin\mathfrak{G}_{R}^{m}}%
\sum_{\boldsymbol{n}}\left\vert \alpha_{\boldsymbol{m,n}}\right\vert ^{2}=0
\]
holds, where
\[
\mathfrak{G}_{R}^{m}:=\left\{  \boldsymbol{m\,}\,;\,\,\frac{\prod_{k=1}%
^{d}m_{k}\,!\,}{m!}\sum_{\boldsymbol{n}}\left\vert \alpha_{\boldsymbol{m,n}%
}\right\vert ^{2}\geq2^{-mR}\right\}
\]
Letting $\boldsymbol{n}^{\ast}=\arg\max_{\boldsymbol{n}}\left\vert
\alpha_{\boldsymbol{m,n}}\right\vert ^{2}$, we have
\begin{align*}
& \sum_{\boldsymbol{m}\notin\mathfrak{G}_{R}^{m}}\sum_{\boldsymbol{n}%
}\left\vert \alpha_{\boldsymbol{m,n}}\right\vert ^{2}\\
& \leq poly\left(  n\right)  \times\max_{\boldsymbol{m}\notin\mathfrak{G}%
_{R}^{m}}2^{-m\left\{  h\left(  \frac{1}{r}\right)  -\sum_{k=1}^{d}\frac
{m_{k}\,}{m}h\left(  \frac{n_{k}^{\ast}}{m_{k}}\right)  +\frac{1}{r}D\left(
\frac{\boldsymbol{n}^{\ast}}{n}||\,\boldsymbol{p}\right)  \right\}  }%
\end{align*}
Observe%
\[
h\left(  \frac{1}{r}\right)  -\sum_{k=1}^{d}\frac{m_{k}\,}{m}h\left(
\frac{n_{k}^{\ast}}{m_{k}}\right)  \geq0
\]
and
\[
\frac{1}{r}D\left(  \frac{\boldsymbol{n}^{\ast}}{n}||\,\boldsymbol{p}\right)
\geq0
\]
holds, and the identity holds if and only if $\boldsymbol{m}=r\boldsymbol{n}%
^{\ast}$ and $\boldsymbol{n}^{\ast}=n\boldsymbol{p}$, respectively. \ Hence,
$R\geq E_{c}\left(  \left\{  \rho_{1}^{m/r,\,m}\right\}  _{m=1}^{\infty
}\right)  $ holds if at least one of the equality does not hold, or
equivalently
\[
\boldsymbol{m}=rn\boldsymbol{p=}m\boldsymbol{p}\in\mathfrak{G}_{R}^{m}%
\]
holds, or equivalently,
\begin{align*}
R  & \geq\frac{1}{m}\log\frac{m!}{\prod_{k=1}^{d}m_{k}\,!}+\frac{-1}{m}%
\log\sum_{\boldsymbol{n}}\left\vert \alpha_{\boldsymbol{m,n}}\right\vert
^{2}\\
& \geq H\left(  \boldsymbol{p}\right)  +h\left(  \frac{1}{r}\right)
-\sum_{k=1}^{d}\frac{m_{k}\,}{m}h\left(  \frac{n_{k}^{\ast}}{m_{k}}\right)
+\frac{1}{r}D\left(  \frac{\boldsymbol{n}^{\ast}}{n}||\,\boldsymbol{p}\right)
+\frac{C}{m}\log m\\
& =H\left(  \boldsymbol{p}\right)  +\frac{C}{m}\log m,
\end{align*}
holds for all $m$. If this holds as $m\rightarrow\infty$, $R\geq E_{c}\left(
\left\{  \rho_{1}^{m/r,m}\right\}  _{m=1}^{\infty}\right)  $. Therefore, we
obtain%
\[
E_{c}\left(  \left\{  \rho_{1}^{m/r,m}\right\}  _{m=1}^{\infty}\right)  \leq
H\left(  \boldsymbol{p}\right)  .
\]

Due to theorem\thinspace\ref{th:Ed-sym}. $R\leq E_{d}\left(  \left\{  \rho
_{1}^{m/r,m}\right\}  _{m=1}^{\infty}\right)  $ holds if and only if%
\[
\lim_{n\rightarrow\infty}\sum_{\boldsymbol{m}\in\mathfrak{G}_{R}^{m}}%
\sum_{\boldsymbol{n}}\left\vert \alpha_{\boldsymbol{m,n}}\right\vert ^{2}=0.
\]
The left hand side can be evaluated in the same way as above,\quad and we can
easily see that the condition is true if
\[
\boldsymbol{m}=rn\boldsymbol{p=}m\boldsymbol{p}\in\mathfrak{G}_{R}^{m}%
\]
holds. This is equivalent to
\[
R\leq H\left(  \boldsymbol{p}\right)  +\frac{C}{m}\log n.
\]
If this holds as $n\rightarrow\infty$, $R\leq E_{d}\left(  \left\{  \rho
_{1}^{m/r,m}\right\}  _{m=1}^{\infty}\right)  $. Therefore, we obtain
\[
H\left(  \boldsymbol{p}\right)  \leq E_{d}\left(  \left\{  \rho_{1}%
^{m/r,m}\right\}  _{m=1}^{\infty}\right)  .
\]
After all, we have
\[
H\left(  \boldsymbol{p}\right)  \leq E_{d}\left(  \left\{  \rho_{1}%
^{m/r,m}\right\}  _{m=1}^{\infty}\right)  \leq E_{c}\left(  \left\{  \rho
_{1}^{m/r,m}\right\}  _{m=1}^{\infty}\right)  \leq H\left(  \boldsymbol{p}%
\right)  .
\]
Therefore, we have the theorem.

\subsection{Proof of theorem\thinspace\ref{th:exponent-dilution-clone}}

We only have to prove (\ref{exponent-dilution-clone}).

To prove "$\geq$", we consider the entanglement dilution protocol mentioned
right before theorem\thinspace\ref{th:Ec-sym}.%

\begin{align}
1-\mathrm{F}^{2^{mR}}\left(  \rho_{1}^{m/r,m}\right)   &  \leq\sum_{\left(
\lambda,l\right)  \notin\mathfrak{F}_{R^{m}}^{m}}b_{\lambda l}^{m}\nonumber\\
&  =\sum_{\left(  \lambda,l\right)  \notin\mathfrak{F}_{R^{m}}^{m}}\sum
_{k=1}^{\dim\mathcal{U}_{\lambda}}c_{\lambda kl}^{m}\nonumber\\
&  \leq\sum_{\left(  \lambda,l\right)  \notin\mathfrak{F}_{R^{m}}^{m}}%
\dim\mathcal{U}_{\lambda}\times\max_{k}c_{\lambda kl}^{m}\nonumber\\
&  \leq poly\left(  n\right)  \times\max_{\left(  \lambda,l\right)
\notin\mathfrak{F}_{R^{m}}^{m}}d_{\lambda}\max_{k}c_{\lambda kl}%
^{m}\nonumber\\
&  \leq poly\left(  n\right)  \times\max_{\lambda:H\left(  \frac
{\boldsymbol{\lambda}}{m}\right)  \geq R-\gamma}d_{\lambda}\max_{k}c_{\lambda
kl}^{m}\,\nonumber\\
&  \leq poly\left(  n\right)  \times\max_{\lambda:H\left(  \frac
{\boldsymbol{\lambda}}{m}\right)  \geq R-\gamma}2^{mH\left(  \frac{\lambda}%
{m}\right)  }\max_{k}c_{\lambda kl}^{m}\,\,,\label{1-F<}%
\end{align}
whee $\gamma$ is an arbitrary positive constant.

Apply random $U_{A}^{\otimes m}\otimes U_{B}^{\otimes m}$ to $\rho_{1}^{n,m}$,
and  denote by $\sigma_{1}^{m}$ the product, which is in the form of
(\ref{random-unitary}). Since this operation is LOCC,
\begin{align}
& 1-\mathrm{F}^{2^{mR}}\left(  \rho_{1}^{m/r,m}\right)  \geq1-\mathrm{F}%
^{2^{mR}}\left(  \sigma_{1}^{m}\right)  \nonumber\\
& =\sum_{\left(  \lambda,l\right)  \notin\mathfrak{F}_{R^{m}}^{m}}b_{\lambda
l}^{m}\geq\sum_{\left(  \lambda,l\right)  \notin\mathfrak{F}_{R}^{m}%
}b_{\lambda l}^{m}\nonumber\\
& \geq\max_{\left(  \lambda,l\right)  \notin\mathfrak{F}_{R}^{m}}d_{\lambda
}b_{\lambda\,l}^{m}\nonumber\\
& \geq\max_{\left(  \lambda,l\right)  \notin\mathfrak{F}_{R}^{m}}d_{\lambda
}\sum_{k}c_{\lambda kl}^{m}\nonumber\\
& \geq\max_{\left(  \lambda,l\right)  \notin\mathfrak{F}_{R}^{m}}d_{\lambda
}\max_{k}c_{\lambda kl}^{m}\nonumber\\
& \geq\max_{\lambda:H\left(  \frac{\boldsymbol{\lambda}}{m}\right)  \geq
R+\gamma^{\prime}}d_{\lambda}\max_{k}c_{\lambda kl}^{m}\nonumber\\
& \geq\frac{1}{poly\left(  m\right)  }\max_{\lambda:H\left(  \frac
{\boldsymbol{\lambda}}{m}\right)  \geq R+\gamma^{\prime}}2^{mH\left(
\frac{\lambda}{m}\right)  }\max_{k}c_{\lambda kl}^{m},\label{1-F>}%
\end{align}
where $\gamma^{\prime}$ is an arbitrary positive constant. Letting
$\gamma\rightarrow0$ and $\gamma^{\prime}\rightarrow0$,  combination of
(\ref{1-F<}) and (\ref{1-F>}) yields%

\[
\frac{-1}{m}\log\left[  1-\mathrm{F}^{2^{mR}}\left(  \rho_{1}^{m/r,m}\right)
\right]  =\min_{\lambda:H\left(  \frac{\boldsymbol{\lambda}}{m}\right)  \geq
R}\min_{k}\left[  \frac{-1}{m}\log c_{\lambda kl}^{m}-H\left(  \frac{\lambda
}{m}\right)  \right]  +o\left(  1\right)  .
\]
A key observation is:
\[
c_{\lambda kl}^{m}=\frac{\prod_{i=1}^{d}\lambda_{i}^{k}\,!}{\lambda^{k}!}%
\sum_{\boldsymbol{\mu}}\,\left\vert \alpha_{\lambda^{k}\boldsymbol{,\,\mu}%
}\right\vert ^{2}%
\]
holds for a $\lambda^{k}$ with $\lambda^{k}\prec\lambda$. This is because: (i)
the eigenvectors are in the form of $\bigotimes_{\kappa}\left\vert i_{\kappa
}\right\rangle $. (ii) the eigenvalue depends only on $m_{j}=\#\left\{
\kappa\,;\,i_{\kappa}=j\right\}  $. Therfore,%
\begin{align}
& \min_{\lambda:H\left(  \frac{\boldsymbol{\lambda}}{m}\right)  \geq R}%
\min_{k}\left[  \frac{-1}{m}\log c_{\lambda kl}^{m}-H\left(  \frac{\lambda}%
{m}\right)  \right]  \nonumber\\
& =\min_{\lambda:H\left(  \frac{\boldsymbol{\lambda}}{m}\right)  \geq R}%
\min_{\lambda^{\prime}:\lambda^{\prime}\prec\lambda}\left[  \frac{-1}{m}%
\log\frac{\prod_{i=1}^{d}\lambda_{i}^{\prime}\,!}{\lambda^{\prime}!}%
\sum_{\boldsymbol{\mu}}\left\vert \alpha_{\lambda^{\prime}\,\boldsymbol{,\mu}%
}\right\vert ^{2}-H\left(  \frac{\lambda}{m}\right)  \right]  \nonumber\\
& =\min_{\lambda:H\left(  \frac{\boldsymbol{\lambda}}{m}\right)  \geq R}%
\min_{\lambda^{\prime}:\lambda^{\prime}\prec\lambda}\min_{\mu:\mu_{i}%
\leq\lambda_{i}^{\prime}}\left[
\begin{array}
[c]{c}%
H\left(  \frac{\lambda\prime}{m}\right)  -H\left(  \frac{\lambda}{m}\right)
+h\left(  \frac{1}{r}\right)  -\sum_{k=1}^{d}\frac{\lambda_{i}^{\prime}\,}%
{m}h\left(  \frac{\mu_{i}}{\lambda_{i}^{\prime}}\right)  \\
+\frac{1}{r}D\left(  \frac{\boldsymbol{\mu}}{n}||\,\boldsymbol{p}\right)
\end{array}
\right]  +o\left(  1\right)  .\label{exponent-tochu}%
\end{align}
Since $H\left(  \cdot\right)  $ is Shur concave,
\[
\min_{\lambda:H\left(  \frac{\boldsymbol{\lambda}}{m}\right)  \geq R}%
\min_{\lambda^{\prime}:\lambda^{\prime}\prec\lambda}\geq\min_{\lambda
,\lambda^{\prime}:H\left(  \frac{\lambda\prime}{m}\right)  \geq H\left(
\frac{\boldsymbol{\lambda}}{m}\right)  \geq R}.
\]
Observe $\lambda$ appears only in \ $-H\left(  \frac{\lambda}{m}\right)  $.
SInce %

\[
-H\left(  \frac{\lambda}{m}\right)  \geq-H\left(  \frac{\lambda^{\prime}}%
{m}\right)  ,
\]
the optimal $\lambda$ equals $\lambda^{\prime}$. Therefore,
(\ref{exponent-tochu}) is lowerbounded by the right hand side of
(\ref{exponent-dilution-clone}) except for $o\left(  1\right)  $-terms. On the
other hand, by simply substituting $\lambda^{\prime}=\lambda$, we can prove
(\ref{exponent-tochu}) is upperbounded by the right hand side of
(\ref{exponent-dilution-clone}).

\section{Output of an optimal cloning machine (2)}

Here, we consider the case where a given state can be an arbitrary pure state.
Our conjecture is that the entanglement cost is again $H\left(  \boldsymbol{p}%
\right)  $. However, we can only show that $H\left(  \boldsymbol{p}\right)  $
is an upperbound.

Letting%

\[
\alpha_{\boldsymbol{\tilde{m},\,\tilde{n}}}:=\sqrt{\frac{\left(  m-n\right)
!\left(  n+d^{2}-1\right)  !}{\left(  m+d^{2}-1\right)  !}}\prod_{k,l=1}%
^{d}\sqrt{\frac{\tilde{m}_{k,l}!}{\tilde{n}_{k,l}!\,\left(  \tilde{m}%
_{k,l}-\tilde{n}_{k,l}\right)  !}\frac{n!}{\prod_{k,l=1}^{d}\tilde{n}%
_{k,l}!\,}p_{k}^{\tilde{n}_{k,k}}\delta_{k,l}},
\]
and $\left\vert R_{\boldsymbol{\tilde{m}}-\boldsymbol{\tilde{n}}}\right\rangle
$ be the internal state of the cloning machine, the final state of optimal
cloning machine is given as follows \cite{FanMatsumoto}.%
\[
\sum_{\boldsymbol{\tilde{m},\,\tilde{l}}}\alpha_{\boldsymbol{\tilde{m}%
,\tilde{n}}}\sqrt{\frac{\prod_{j,k=1}^{d}\tilde{m}_{j,k}!}{m!}}\sum
_{\#\left\{  \left(  \kappa,\mu\right)  :\left(  i_{\kappa},\,i_{\mu}\right)
=\left(  j,\,k\right)  \right\}  =\tilde{m}_{j,k}}\bigotimes_{\kappa,\mu
=1}^{m}\left\vert i_{\kappa}\right\rangle \left\vert i_{\mu}\right\rangle
\otimes\left\vert R_{\boldsymbol{\tilde{m}}-\boldsymbol{\tilde{n}}%
}\right\rangle .
\]
Denote by $\rho_{2}^{n,m}$ the state after tracing out the internal state of
cloning machine. $\rho_{2}^{n,m}$ is probability mixture of
\[
\frac{1}{\sqrt{\beta_{\boldsymbol{\tilde{n}}}}}\sum_{\boldsymbol{\tilde{m}}%
}\alpha_{\boldsymbol{\tilde{m},\,\tilde{n}}}\sqrt{\frac{\prod_{j,k=1}%
^{d}\tilde{m}_{j,k}!}{m!}}\sum_{\#\left\{  \left(  \kappa,\mu\right)  :\left(
i_{\kappa},\,i_{\mu}\right)  =\left(  j,\,k\right)  \right\}  =\tilde{m}%
_{j,k}}\bigotimes_{\kappa,\mu=1}^{m}\left\vert i_{\kappa}\right\rangle
\left\vert i_{\mu}\right\rangle ,
\]
with the probability $\beta_{\boldsymbol{\tilde{n}}}$, where
\[
\beta_{\boldsymbol{\tilde{l}}}=\sum_{\boldsymbol{\tilde{m}}}\left\vert
\alpha_{\boldsymbol{\tilde{m},\tilde{n}}}\right\vert ^{2}.
\]

Now we apply
\[
\bigotimes_{\kappa=1}^{m}\sum_{i,j=1}^{d}e^{\sqrt{-1}\omega_{i,\kappa}^{A}%
}\left\vert i\right\rangle \left\langle i\right\vert \otimes e^{\sqrt
{-1}\omega_{j,\kappa}^{B}}\left\vert j\right\rangle \left\langle j\right\vert
,
\]
where $\omega_{i,\kappa}^{A}$, $\omega_{j,\kappa}^{B}$ are chosen
independently randomly. After the application of this operation, Bob's local
view will have the density matrix with the eigenvector $\bigotimes_{\kappa
=1}^{m}\left\vert i_{\kappa}\right\rangle $ and the corresponding eigenvalue
\[
\sum_{\sum_{k=1}^{d}\tilde{m}_{j,k}=m_{j}^{A}}\frac{\prod_{j,k=1}^{d}\tilde
{m}_{j,k}!}{m!}\left\vert \alpha_{\boldsymbol{\tilde{m},\tilde{n}}%
}\,\right\vert ^{2}\,,
\]
where $m_{j}^{A}=\#\left\{  \kappa;i_{\kappa}=j\right\}  $. Each eigenvalue
has $m!/\prod_{j=1}^{d}m_{j}^{A}!$ folds degeneracy. To compute these, it is
easier to apply dephasing at both parties first, and take partial trace
later.\bigskip

\begin{lemma}
Let $\boldsymbol{q}^{n}$ be the spectrum of the reduced density matrix
$\rho^{n}$. Then,
\[
\mathrm{p-}\varlimsup_{n\rightarrow\infty}\frac{-1}{n}\log q_{i}^{n}%
\]
is decreasing by application of the above operation.
\end{lemma}

\begin{proof}
Consider a pure state $\left\vert \psi^{n}\right\rangle \,$\ with the Schmidt
coefficients $\boldsymbol{q}^{n}$. Then, $E_{c}\left(  \left\{  \left\vert
\psi^{n}\right\rangle \,\right\}  _{n=1}^{\infty}\right)  =\mathrm{p-}%
\varlimsup_{n\rightarrow\infty}\frac{-1}{n}\log q_{i}^{n}$. Therefore,
$\mathrm{p-}\varlimsup_{n\rightarrow\infty}\frac{-1}{n}\log q_{i}^{n}$ has to
be Shur concave, and should be monotone with respect to the probabilistic unitary.
\end{proof}

\begin{remark}
Since the above dephasing operation is LOCC, the entanglement cost of the
resultant state is a lowerboud of it of the optimal clone. However, this state
is not supported on the symmetric subspace anymore, and we cannot apply our formula.
\end{remark}

Hence, letting
\[
\mathfrak{H}_{R}^{n}=\left\{  \left(  \boldsymbol{\tilde{m}}^{A}%
,\boldsymbol{\tilde{n}}\right)  \,;\,\sum_{\boldsymbol{\tilde{m}}:\sum
_{k=1}^{d}\tilde{m}_{j,k}=m_{j}^{A}}\frac{\prod_{j,k=1}^{d}\tilde{m}_{j,k}%
!}{m!}\left\vert \alpha_{\boldsymbol{\tilde{m},\tilde{n}}}\,\right\vert
^{2}\geq2^{-nR}\right\}  ,
\]
and denoting by $\boldsymbol{\tilde{p}}$ the probability distribution
$p_{i}\delta_{i,j}$ over the set $\{(i,j);i,j=1,\cdots d\}$, $R\geq
E_{c}\left(  \left\{  \rho_{2}^{m/r\,,\,m}\right\}  _{m=1}^{\infty}\right)  $
holds is the following sum goes to 0.
\begin{align}
& \sum_{\left(  \boldsymbol{\tilde{m}}^{A},\boldsymbol{\tilde{n}}\right)
\notin\mathfrak{h}_{R}^{n}}\frac{m!}{\prod_{j=1}^{d}m_{j}^{A}!}\sum
_{\sum_{k=1}^{d}\tilde{m}_{j,k}=m_{j}^{A}}\frac{\prod_{j,k=1}^{d}\tilde
{m}_{j,k}!}{m!}\left\vert \alpha_{\boldsymbol{\tilde{m},\tilde{n}}%
}\,\right\vert ^{2}\label{tail-upper}\\
& \leq poly\left(  m\right)  \times\max_{\boldsymbol{\tilde{m}}^{A}%
,\boldsymbol{\tilde{n}},\boldsymbol{\tilde{m}}}2^{-m\left\{  -H\left(
\frac{\boldsymbol{m}^{A}}{m}\right)  +H\left(  \frac{\boldsymbol{\tilde{m}}%
}{m}\right)  +h\left(  \frac{1}{r}\right)  -\sum_{k,l}\frac{\tilde{m}_{k,l}%
}{m}h\left(  \frac{\tilde{n}_{k,l}}{\tilde{m}_{k,l}}\right)  +\frac{1}%
{r}D\left(  \frac{\boldsymbol{\tilde{n}}}{n}||\,\boldsymbol{\tilde{p}}\right)
\right\}  },\nonumber
\end{align}
where the maximization is taken over all $\boldsymbol{\tilde{m}}%
^{A},\boldsymbol{\tilde{n}},\boldsymbol{\tilde{m}}$ with%
\begin{equation}
\left(  \boldsymbol{\tilde{m}}^{A},\boldsymbol{\tilde{n}}\right)
\notin\mathfrak{h}_{R}^{n},\,\tilde{n}_{jk}\leq\tilde{m}_{jk},\boldsymbol{\,}%
\,\sum_{k=1}^{d}\tilde{m}_{j,k}=m_{j}^{A}.\label{range-variables}%
\end{equation}
Observe that
\begin{align*}
-H\left(  \frac{\boldsymbol{m}^{A}}{m}\right)  +H\left(  \frac
{\boldsymbol{\tilde{m}}}{m}\right)   & \geq0,\\
h\left(  \frac{1}{r}\right)  -\sum_{k,l}\frac{\tilde{m}_{k,l}}{m}h\left(
\frac{\tilde{n}_{k,l}}{\tilde{m}_{k,l}}\right)   & \geq0,\\
\frac{1}{r}D\left(  \frac{\boldsymbol{\tilde{n}}}{n}||\,\boldsymbol{\tilde{p}%
}\right)   & \geq0.
\end{align*}
The identity in each inequality holds if and only if \
\begin{align*}
\frac{\boldsymbol{m}^{A}}{m}  & =\frac{\boldsymbol{\tilde{m}}}{m},\\
\frac{\boldsymbol{\tilde{n}}}{n}  & =\frac{\boldsymbol{\tilde{m}}}{m},\\
\boldsymbol{\tilde{p}}  & \boldsymbol{=}\frac{\boldsymbol{\tilde{n}}}{n},
\end{align*}
holds, respectively. Hence, the right hand side of (\ref{tail-upper})
converges to 0 if one of these does not hold, or equivalently,
\[
\boldsymbol{\tilde{m}}=r\boldsymbol{\tilde{n}}=m\boldsymbol{\tilde{p}}%
\in\mathfrak{h}_{R}^{n},
\]
or equivalently,%

\begin{align*}
R  & \geq H\left(  \boldsymbol{\tilde{p}}\right)  +h\left(  \frac{1}%
{r}\right)  -\sum_{k,l}\frac{\tilde{m}_{k,l}}{m}h\left(  \frac{1}{r}\right)
+\frac{1}{r}D\left(  \boldsymbol{\tilde{p}}||\,\boldsymbol{\tilde{p}}\right)
+\frac{C}{m}\log m\\
& =H\left(  \boldsymbol{\tilde{p}}\right)  +\frac{C}{m}\log m.
\end{align*}
If this holds as $m\rightarrow\infty$, \ $E_{c}\left(  \left\{  \rho
_{2}^{m/r\,,\,m}\right\}  _{m=1}^{\infty}\right)  \leq R$. Therefore, we have
\[
E_{c}\left(  \left\{  \rho_{2}^{m/r\,,\,m}\right\}  _{m=1}^{\infty}\right)
\leq H\left(  \boldsymbol{\tilde{p}}\right)  =H\left(  \boldsymbol{p}\right)
.
\]

\section{Discussions}

We first computed entanglement cost and distillable entanglement of non-i.i.d
mixed state explicitly, and also gave general formula.  We also have shown
that universal entanglement concentration can be extended to arbitrary
symmetric states. 

Surprisingly, the real $m$ copies and optimal clone under some assumption are
the same in entanglement quantities. \ This is rather surprising since
$F\left(  \rho_{1}^{m/r,m},\left\vert \phi\right\rangle ^{\otimes m}\right)
\approx r^{d}$, and these two states are not so close. However, with closer
look, entanglement of $\rho_{1}^{m/r,m}$ and $\left\vert \phi\right\rangle
^{\otimes m}$ are somewhat different. More concretely, they differ in error
exponent of entanglement dilution and distillation. This motivate to use
entanglement cost and distillable entanglement with restriction to error
exponent. Such a measure had been closely studied in \cite{Morikoshi} for
i.i.d. pure ensembles, and in  \cite{Hayashi:2006} for general purestates.
However, detailed analysis for mixed state ensembles, either i.i.d. or
non-i.i.d, are still to be studied. 

Another interesting open problem is  the entanglement cost and distillable
entanglement of optimal clone of totally unknown purestates. Are they also
same as these of $\left\vert \phi\right\rangle ^{\otimes m}$?

\appendix

\section{Group representation theory}

\label{appendixA}

\begin{lemma}
\label{lem:decohere} Let $U_{g}$ and $U_{g}^{\prime}$ be an irreducible
representation of $G$ on the finite-dimensional space $\mathcal{H}$ and
$\mathcal{H}^{\prime}$, respectively. We further assume that $U_{g}$ and
$U_{g}^{\prime}$ are not equivalent. If a linear operator $A$ in
$\mathcal{H}\oplus\mathcal{H}^{\prime}$ is invariant by the transform
$A\rightarrow U_{g}\oplus U_{g}^{\prime}AU_{g}^{\ast}\oplus U_{g}^{^{\prime
}\ast}$ for any $g$, $\mathcal{H}A\mathcal{H^{\prime}}=0$ ~\cite{GW}.
\end{lemma}

\begin{lemma}
\label{lem:shur} (Shur's lemma~\cite{GW}) Let $U_{g}$ be as defined in
lemma~\ref{lem:decohere}. If a linear map $A$ in $\mathcal{H}$ is invariant by
the transform $A\rightarrow U_{g}AU_{g}^{\ast}$ for any $g$,
$A=c\boldsymbol{1}_{\mathcal{H}}$.
\end{lemma}

\section{Representation of symmetric group and SU($d$)}

Due to \cite{GW}, we have%

\begin{align}
\dim\mathcal{U}_{\lambda}  &  =\frac{\prod_{i<j}\left(  l_{i}-l_{j}\right)
}{\prod_{i=1}^{d-1}\left(  d-i\right)  !},\label{dim-representation-1}\\
d_{\lambda}  &  =\dim\mathcal{V}_{\lambda}=\frac{n!}{\prod_{i=1}^{d}\left(
\lambda_{i}+d-i\right)  !}\prod_{i<j}\left(  l_{i}-l_{j}\right)
,\label{dim-representation-2}%
\end{align}
with $l_{i}:=\lambda_{i}+d-i$. \ It is easy to show%
\begin{equation}
\log\dim\mathcal{U}_{\lambda}\leq d^{2}\log n.\label{dim-zero-rate}%
\end{equation}

Let $a_{\lambda}^{\phi}=\mathrm{Tr}\left\{  \mathcal{W}_{\lambda,A}\left(
\mathrm{Tr}_{B}|\phi\rangle\langle\phi|\right)  ^{\otimes n}\right\}  $ and
the formulas in the appendix of \cite{Ha} says%

\begin{align}
\left\vert \frac{\log d_{\lambda}}{n}-\mathrm{H}\left(  \frac{\lambda}%
{n}\right)  \right\vert  &  \leq\frac{d^{2}+2d}{2n}\log
(n+d),\label{grep-type-1}\\
\sum_{\frac{\lambda}{n}\in\mathfrak{R}}a_{\lambda}^{\phi} &  \leq\left(
n+1\right)  ^{d\left(  d+1\right)  /2}\exp\left\{  -n\min_{\boldsymbol{q}%
\,\in\mathfrak{R}}\mathrm{D}\left(  \boldsymbol{q}||\boldsymbol{p}\right)
\right\}  ,\label{grep-type-2}%
\end{align}
where $\mathfrak{R}$ is an arbitrary closed subset.

\section{}
\end{document}